\documentclass[12pt]{article}
\usepackage{amssymb}
\usepackage{epsf}
\begin{document}
\baselineskip8mm
\title{Dynamics of Bianchi IX universe with massive scalar field}
\author{A. V. Toporensky$^{1}$\footnote{Electronic mail:  lesha@sai.msu.su}
$\;$ and V. O. Ustiansky$^{1}$\footnote{Electronic mail: ustiansk@sai.msu.su}}
\date{}
\maketitle
\hspace{-6mm}
$^{1}${\small \it  Sternberg Astronomical Institute, Moscow University, Moscow,
119899, Russia}\\
\\
\begin{abstract}
The dynamics of the Bianchi IX cosmological model with minimally coupled
massive real scalar field is studied.
The possibility of non-singular transition from contraction to expansion is
shown.
A set of initial conditions that lead to non-singular solutions is studied
numerically.
\end{abstract}
\thispagestyle{empty}
\newpage
\section{Introduction}
During last years the dynamics of closed FRW universe with massive minimally
coupled real scalar field has become a subject of investigations. Being one
of the simplest cosmological models it has a peculiar property. Namely the
universe of this type can make a non-singular transition from contraction to
expansion (in other worlds a scale factor can posses a local minimum). Such
a transition is often referred to as 'bounce'.

It turns out that a universe in principle can make an arbitrary number of
bounces. Page \cite{Page} was the first who conjectured the presence of a
fractal set of infinitely bouncing solutions, later Cornish
and Shellard \cite{CornSh} proved that this system is really chaotic. In
later works \cite{simple}--\cite{entropy} , \cite{exp} the chaotic properties
of this and more complicated models were studied.

Numerical investigations have shown that the initial conditions space has a
quasi-periodical structure : there are narrow zones which lead to bouncing
solutions separated by wide zones which lead to singular solutions. These
investigations also have shown that in former zones the solutions exhibits
a sensible dependence of initial conditions -- a property peculiar
 to the chaotic
dynamical systems. Moreover, starting from one of such zones a solution may pass 
after bounce through another zone of this type (or through the same one)
and suffer a new bounce. After each bounce the situation is
similar to the previous one. This lead to the existence of a fractal set of
unstable infinitely bouncing solutions. In the dynamical chaos theory
such set is called as a strange repellor -- a structure which can be
found
with in many chaotic dynamical systems without dissipation. The presence of this
fractal structure of solutions corresponds to the fractal structure of sets
of initial conditions leading to bouncing solutions~\cite{CornSh}.

But the Universe at the earliest stages of its
evolution have not necessary had a metric close to FRW one. So for
studying the Universe at this time one often invoke more general
metrics. The simplest non--FRW metric with positive curvature is the
Kantowski--Sachs one.  However this case does not include the closed
FRW metric, and moreover there are no bouncing solutions in this case
(see appendix). More complicated metric is Bianchi IX model. This is
 the only model that includes closed FRW metric \cite{MTW}. So in this
 paper we investigate the Bianchi IX universe with massive scalar
field.

The Bianchi IX universe is well known due to its mixmaster approach to
singularity \cite{MTW}. It worth to notice that mixmaster
chaos \cite{CornLev,CornLev2} principally differs from the studying one as in
the case of mixmaster it is the evolution of anisotropic variables which
exhibit chaotic properties while the overall volume of the universe decrease
monotonically. In the considered case  it is the evolution
of the total volume of the universe which exhibit chaotic properties.

\section{Equations of motion}
Let us consider a Bianchi IX universe with minimally coupled massive scalar
field with the action as follows (we use units $c=\hbar=1$, $G=1/{8\pi}$):
\begin{equation}
S=\int\left(-\frac{R}{2}+\frac{1}{2} \varphi_{,i}\varphi^{,i}-\frac{m^2}{2}
\varphi^2\right)\sqrt{-g}\; d^4 x\; .
\end{equation}

\noindent The metrics has the following form :
\begin{equation}
ds^2=dt^2-\eta_{ \alpha \beta}(t)\omega^\alpha \omega^\beta \;,
\end{equation}

\noindent where
\begin{equation}
\eta_{ \alpha \beta}(t)=\left(
\begin{array}{ccc}
a(t)^2&0&0\\
0&b(t)^2&0\\
0&0&c(t)^2
\end{array}
\right)\;,
\end{equation}

\noindent and $\omega^\alpha$ are differential 1--forms invariant under SO(3)
transformations.

Assuming $\varphi$ to be a function of time variable only one can write
the Einstein's field equations in following form :
\begin{equation}
\frac{(\dot a bc)\dot {}}{abc}+\frac{1}{2a^2b^2c^2}[a^4-(b^2-c^2)^2]=
\frac{m^2}{4}\varphi^2\;,\label{BKL1}
\end{equation}
\begin{equation}
\frac{(a\dot b c)\dot{}}{abc}+\frac{1}{2a^2b^2c^2}[b^4-(a^2-c^2)^2]=
\frac{m^2}{4}\varphi^2\;,
\end{equation}
\begin{equation}
\frac{(ab\dot c)\dot{}}{abc}+\frac{1}{2a^2b^2c^2}[c^4-(a^2-b^2)^2]=
\frac{m^2}{4}\varphi^2\;,
\end{equation}
\begin{equation}
\ddot\varphi+\left(\frac{\dot a}{a}+\frac{\dot b}{b}+\frac{\dot c}{c}
\right)\dot\varphi+m^2\varphi=0 \label{matt1}
\end{equation}

\noindent with the first integral of motion :
\begin{equation}
\frac{1}{2} \dot\varphi^2+\frac{m^2}{2}\varphi^2=2\left(\frac{\dot
a}{a}\frac{\dot
 b}{b}+\frac{\dot a}{a}\frac{\dot b}{b}+\frac{\dot b}{b} \frac{\dot
c}{c}\right)
+\frac{1}{a^2}+\frac{1}{b^2}+\frac{1}{c^2}-\frac{a^4+b^4+c^4}{2a^2
b^2 c^2}\;,\label{cnstr1}
\end{equation}

\noindent where $\quad \dot {} \equiv \frac{d}{dt}$ .

In our investigation it is more convenient to work with ADM variables
$\left\{\Omega(\tau), \beta_+(\tau), \beta_-(\tau)\right\}$ which are
functions of a new
time variable $\tau$ : $dt=m^{5} e^{3\Omega}d\tau$ . Relations between ADM
variables and the old ones are :
\begin{equation}
\left( \begin{array}{c}
\Omega\\
\beta_+\\
\beta_-
\end{array}\right)=
\left(\begin{array}{rrr}
\frac{1}{3} &\frac{1}{3} &\frac{1}{3} \\
\frac{1}{6}&\frac{1}{6}&-\frac{1}{3}\\
\frac{\sqrt{3}}{6}&-\frac{\sqrt{3}}{6}&0
\end{array}\right)
\left(\begin{array}{c}
\ln a\\
\ln b\\
\ln c
\end{array}\right)
-\left(\begin{array}{c}
\ln m \\ 0 \\ 0 \end{array}\right)\;.\label{conv}
\end{equation}
It is easily seen from (\ref{conv}) that variable $\Omega$ describes the
volume of the universe while $\beta_\pm$ describes it's anisotropy.

The considered dynamical system admits Hamiltonian formalism. In terms of
ADM variables the Hamiltonian has the following form :
\begin{equation}
H=\frac{1}{2}(-p_\Omega^2+p_+^2+p_-^2+\frac{1}{12}p_\varphi^2)
+\frac{1}{8}e^{4\Omega} V(\beta_+,\beta_-)+\frac{1}{24}\varphi^2
e^{6\Omega}\;,
\end{equation}
where
\begin{equation}
V(\beta_+, \beta_-)=\frac{1}{3}(e^{-8\beta_+}+2e^{4\beta_+}(\cosh 4\sqrt{3}
\beta_- -1)-4e^{-2\beta_+}\cosh 2\sqrt{3}\beta_-)\;.\label{defV}
\end{equation}
The  Hamiltonian equations leads to the following :
\begin{eqnarray}
\Omega'' &=&\frac{1}{2}e^{4\Omega}V(\beta_+, \beta_-)
           +\frac{1}{4}\varphi^2 e^{6\Omega},\label{ADM1}\\
\beta_+''&=&\frac{1}{3}e^{4\Omega}\left[ e^{-8\beta_+}-e^{4\beta_+}(\cosh 4
            \sqrt{3}\beta_- -1)-e^{-2\beta_+}\cosh 2\sqrt{3}\beta_-\right],\\
\beta_-''&=&\frac{\sqrt{3}}{3} e^{4\Omega}\left[ e^{-2\beta_+}\sinh 2
             \sqrt{3}\beta_- -e^{4\beta_+}\sinh 4\sqrt{3}\beta_-\right],\\
\varphi''&=&-\varphi e^{6\Omega}\label{matt2}
\end{eqnarray}
and the constraint equation $H=0$ takes the form :
\begin{equation}
-\Omega'^2+\beta_+'^2+\beta_-'^2+\frac{1}{12}\varphi'^2+\frac{1}{4}e^{4\Omega}
V(\beta_+, \beta_-)+\frac{1}{12}\varphi^2 e^{6\Omega}=0 \;,\label{constr}
\end{equation}
where $\quad ' \equiv \frac{d}{d\tau}$ .

To make further investigations we need to generalize the notion of bounce to
Bianchi IX case. Being defined as a non-singular transition from contraction
to expansion of the universe bounces thus corresponds to local minima of
$\Omega$. That is why ADM variables is more preferable choice
for our purposes.

Substituting $\Omega'=0$ into
constraint equation (\ref{constr}) and combining it with the condition
$\Omega''>0$ it is easily seen that bounces can occur only if :
\begin{eqnarray}
V(\beta_+, \beta_-)+\frac{1}{3}\varphi^2 e^{2\Omega}\le0\;,\label{pb1} \\
V(\beta_+, \beta_-)+\frac{1}{2}\varphi^2 e^{2\Omega}>0\;. \label{pb2}
\end{eqnarray}
The above inequalities at least implies $V(\beta_+, \beta_-)<0$ which can be satisfied
only in a narrow region around the isotropic case $(\beta_+=\beta_-=0)$ where
$V(\beta_+, \beta_-)$ suffer it's minimum. This region is represented on
fig.~1 where the levels $V(\beta_+, \beta_-)=const$ are shown for
the values of $const=-.99$(inner curve), $const=-.90$ , $const=-.75$ ,
$const=-.50$ and $const=0$(bold lines). The potential  has a minimum in
point $\beta_+=\beta_-=0$ (isotropic case) and $V(0,0)=-1$. It also follows
from (\ref{constr}) that if a bounce occur the possible values of
$\beta_\pm '$ must satisfy criterion : $\beta_+'^2+\beta_-'^2<1$ as
$V(\beta_+, \beta_-)\ge -1$ for all values of $\beta_\pm$ .

\begin{figure}[h]
\epsfxsize=0.65\hsize 
\epsfysize=0.65\hsize
\centerline{\epsfbox{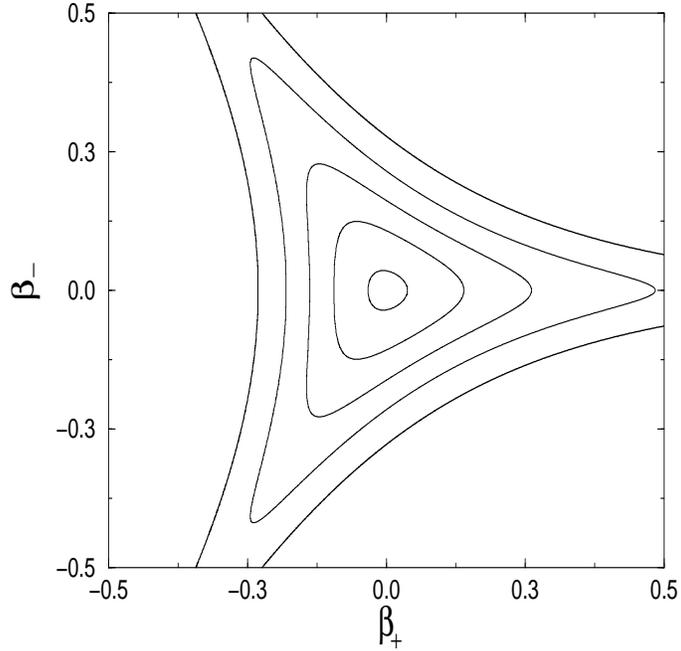}}
\label{Veqv}
\caption{A levels of $V(\beta_+, \beta_-)=const$.}
\end{figure}

Before starting the next section it should be mentioned that Belinsky and
Khalatnikov \cite{massless} were the first who studied the Bianchi IX model
whith scalar field. Thay considered only massless case and found that the
presence of the scalar field of this type destroys the Mixmaster
oscillatory regime when the universe collapses to singularity. This results
holds in massive case as an asymptotic regime.

\section{Numerical results}
The equations of motion (\ref{ADM1})--(\ref{matt2}) can be solved
numerically. In order to obtain numerical solution one need to fix eight
initial conditions. The constraint equation allows to express one of them in
terms of the others (in our work we determined $\dot\varphi(0)$ in this way).
Moreover as the considered system has at least one maximal expansion point
(i.e a point where $\Omega$ suffer (local) maximum), one can fix
$\Omega'(0)=0$ without loss of generality.

For simplicity in our calculations
we always fixed $\varphi(0)=0$ . So we studied the
dependence of numerical solution from five initial conditions : $\Omega(0)$,
$\beta_+(0)$, $\beta_-(0)$, $\beta_+'(0)$, $\beta_-'(0)$ .

The equations of motion have been solved numerically using Bulirsch--Stoer
method \cite{numrecip}. As equations (\ref{ADM1})--(\ref{matt2}) requires
evaluations of transcendental functions while equations (\ref{BKL1})--
(\ref{matt1}) requires only arithmetical operations we performed calculations
in variables $\left\{ a, b, c \right\}$ in order to speed-up them while the
initial conditions and results were expressed in terms of $\left\{ \Omega,
\beta_+, \beta_-\right\}$ . The constraint equation (\ref{cnstr1}) has been
used to check the accuracy.

First of all we tested our routines on already known fridmanian case. On
fig.~2 we presented a first four closest to singularity
simplest symmetric periodical solutions in a projection onto $(\varphi,
\Omega)$ plane. Term 'simplest' means that all of the maximal expansion
points for given solution coincide in this plane and symmetric means that
these points lies on axis $\varphi=0$. The values of $\Omega$ in the
above mentioned points are : $\Omega_1=0.160832$ (short dashed line),
$\Omega_2=0.760135$ (long dashed line),
$\Omega_3=1.141779$
(dot-dashed line), $\Omega_4=1.418301$ (solid line). Near each of
these values $\Omega_i$ there are narrow intervals which lead to bouncing
solutions.

\begin{figure}[h]
\epsfxsize=0.55\hsize 
\epsfysize=0.55\hsize
\centerline{\epsfbox{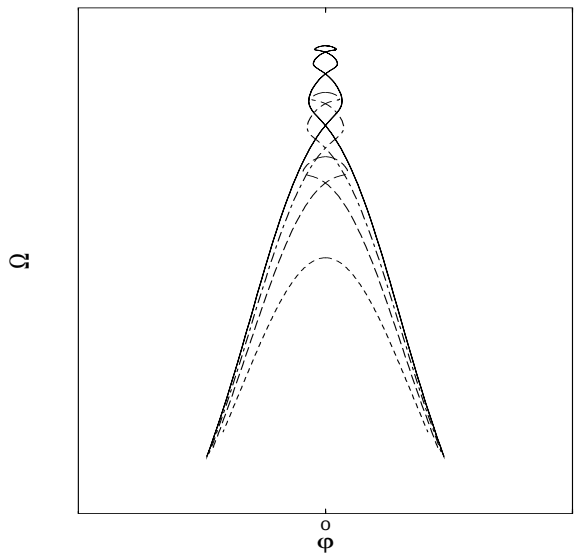}}
\label{periodic}
\caption{Examples of periodical solutions in Fridmanian case.}
\end{figure}

As $\Omega'(0)=0$ the initial values $\beta_\pm(0)$ and $\beta_\pm'(0)$ have
similar restrictions as in the case of bounce with the only exception that
(\ref{pb2}) is no longer required. Numerical investigations have shown the
presence of bouncing solutions only in the case of sufficiently weak
anisotropy near the corresponding fridmanian bouncing solutions. These 
investigations  have also shown that for some ranges of
initial values of
$\Omega$, $\beta_\pm$ and $\beta_\pm'$ solutions exhibits sensible
dependence of initial conditions.

To illustrate the structure of the sets of initial conditions which leads to
bouncing solutions we calculated a 2d slices through the initial conditions
space. For given value of $\Omega(0)$ each slice represents a $500\times
500$ grid in a plane of two of the anisotropy variables while the rest
of them equals to zero. For each knot of the grid it has been
calculated a numerical solution on a sufficiently large interval of $t\in
[0, 50]$ . If a maximal value of  $\Omega$ on this solution after bounce 
$\Omega_{max}$ is grater then $\Omega_4$ we encoded corresponding dot in a
black color. If $\Omega_4\ge\Omega_{max}>\Omega_3$ 
we encoded it in a dim gray color. If $\Omega_3\ge\Omega_{max}>
\Omega_2$ we encoded it in a gray color and if $\Omega_2\ge\Omega_{max}>
\Omega_1$ we used a light gray color.  
We considered only solutions which
satisfies criterion $\Omega_{max}\ge\Omega_1$ as they are of the most interest.
We do not consider the trajectories with the next maximum
expantion point lying closer to the singularity than the first bounce
interval, becouse in this case the trajectory have only a zigzag,
which can not change the fate of the trajectory falling into the
singularity.

We calculated such slices for subsequent
values of $\Omega$ with sufficiently small steps so that they can be
viewed as a 'movie'. Such movies can be found at
"http://www.xray.sai.msu.su/\~{}ustiansk/sciwork/gr-qc/001.html".
Examples of these slices are shown in fig.3--4.

Numerical calculations have shown that there are narrow intervals of initial
values $\Omega$ that lead to bouncing solutions separated by wide intervals
that lead to singular solutions in full analogy with the fridmanian case. For
the closest to singularity interval (an interval which contain
$\Omega_1$) in a $(\beta_+, \beta_-)$ plane one can observe the picture that
follows. For the values $\Omega\lesssim 0.1549$ there are no bouncing solutions.
With the increase of $\Omega$ bouncing solutions appears in a small region
around isotropic case which grow with the growth of $\Omega$ up to the value
$\Omega\approx 0.1567$ when a fraction of considered anisotropic bouncing
solutions begins to decrease until they vanishes at $\Omega\approx 0.1608$ .

In the case of $(\beta_+, \beta_+')$ and $(\beta_-, \beta_-')$ slices the
general picture slightly differs. With $\Omega$ increasing from the value
$\Omega \approx 0.1525$ anisotropic bouncing solutions appears and invoke
isotropic case at already mentioned value $\Omega\approx 0.1550$. Thus in
considered case the interval of $\Omega$ which allows bouncing solutions
becomes slightly wider than in fridmanian one.

For the interval which contain $\Omega_2$ the picture is following. For both
studied planes $(\beta_+, \beta_-)$ and $(\beta_-, \beta_-')$ bouncing
solutions appears when $\Omega\approx 0.7593$ in a small region around
isotropic case. With the growth of $\Omega$ this region grows until
$\Omega\approx 0.7603$ . After that the
value of $\Omega_{max}$ becomes smaller than $\Omega_1$ for 
initially anysotropy close to zero so such solutions were discarded 
and thus the region
have a 'ring' shape. For the values $\Omega\approx 0.762$ this regions
vanishes.  So in considered case the interval of $\Omega$ is wider than
in fridmanian one. The similar picture is valid for the interval that
contain $\Omega_3$.

On several slices one can easily see a narrow 'rings' (one of them is
pointed by two arrows in fig.4) inside the region of
initial conditions that lead to bounces. The existence of these 'rings'
is caused by the fact that solutions which correspond to these 'rings'
suffers bounce and the next point of maximal expansion also occurs inside
one of the regions that leads to bounce. So such solutions suffer at least
two bounces. These 'rings' forms a fractal structure. The example of such
structure is shown in fig.5.

For the splits in $(\beta_+, \beta_-)$ it is possible to determine the 
mesure of bouncing solutions in a simple way if we start from $\varphi(0)=0$.
In this case the measure of all possible
initial values of $\beta_{\pm}$ , being defined as the area bounded by the
bold lines in fig.1 , equals $\sim 0.48$. The maximal fractions of initial
values of $\beta_{\pm}$ that lead to acceptable bouncing solutions to all
posible initial values for first three subsequent intervales of $\Omega$
are approximately $0.0026$, $0.00041$, $0.00012$.

\section{Conclusions}
In this paper we studed the Bianki IX cosmological model with the massive
scalar field. In particular the influence of initial shear on the
possibility of transition from contraction to expansion was investigated.
We show that only a very restricted  set of initial conditions with
shear can led to the bouncing trajectories. Their mesure with respect to
all physically admissible initial conditions for the first 3 zones
are restricted by $2.6 \times 10^{-3}$, $4.1 \times 10^{-4}$ and
$1.2 \times 10^{-4}$ respectively. Corresponding areas of the initial
conditions were found
numerically also for other cross-sections of the initial conditions space.
In all cases, bounces are possible for a ruther narrow zones near
corresponding friedmannian trajectory.

On the other hand, the structure of chaos is similar to known in
the Friedmann case:
there are narrow intervals of initial
values $\Omega$ that lead to bouncing solutions separated by wide intervals
that lead to singular solutions. The width of the bounce intervals
may be enlarged up to $3$ times in comparition with the Friedmann case
for the several closest to the singularity intervals, but the
qualitative picture is the same one.

\section*{Appendix. No chaos in Kantowski--Sachs universe with a massive
scalar field}

The equations of motion for the Kantowski--Sachs universe filled by a scalar
field are
\begin{equation}
\frac{\ddot b}{b} - \frac{\dot a \dot b}{a b} + \frac{\dot \phi^2}{2} = 0,
\end{equation}
\begin{equation}
\frac{\ddot b}{b} + \frac{\dot a \dot b}{a b} + \frac {\ddot a}{a} + \frac{\dot
\phi^2}{2}=V(\phi),
\end{equation}
\begin{equation}
\ddot \phi + (\frac{\dot a}{a} + 2 \frac{\dot b}{b})\dot \phi + V'(\phi)=0,
\end{equation}
with the constraint
\begin{equation}
2 \frac{\dot a \dot b}{a b} + \frac{\dot b^2}{b^2} + \frac{1}{b^2} =
\frac{\dot \phi^2}{2} + V(\phi)
\end{equation}
where $a, b$ - the scale factors, $V(\phi)$ - a potential of the scalar
field, and the prime indicates the derivative with respect to $\phi$.

It is easy to see from the Eq.(19) that in all the possible turning points
with respect to $b$ (i.e. the points in which
$\dot b =0$), the corresponding acceleration
\begin{equation}
\ddot b = -b \dot \phi^2/2
\end{equation}
is always nonpositive, so we can have only the points of maximal
expansion for the scale factor $b$.

Combining Eqs.(19 -- 20) we receive, on the other side, that the acceleration
in the turning points with respect to $a$ is
\begin{equation}
\ddot a = a V(\phi),
\end{equation}
which is nonnegative for an arbitrary nonnegative scalar field potential
$V(\phi)$, and we have only the points of minimal contraction for the
scale factor $a$.

As a result, we can claim from Eqs.(23 -- 24) that the type of chaos
associated with the oscillation of the scale factors is absent in
Kantowski--Sachs universe filled by the scalar field with an arbitrary
nonnegative potential.

\baselineskip8mm

\newpage
{\large \bf Figure captions}

Fig.3: A slice through initial condition space with fixed $\Omega=0.1573$ which
represents a $500\times 500$ grid in $(\beta_+, \beta_-)$ plane.
For each knot of the grid it has been calculated a numerical solution.
If a maximal value of  $\Omega$ on this solution $\Omega_{max}>
\Omega_4$ we encoded corresponding dot in a black color. If
$\Omega_4\ge\Omega_{max}>\Omega_3$
we encoded it in a dim gray color. If $\Omega_3\ge\Omega_{max}>
\Omega_2$ we encoded it in a gray color and if $\Omega_2\ge\Omega_{max}>
\Omega_1$ we used a light gray color.

Fig.4: A slice analogous to previous one for the value $\Omega=0.7604$.
One of thing rings which represents second order structure is pointed by two
arrows. These rings are formed by trajectories which suffer two bounces.

Fig.5: Example of fractal structure in $(\beta_+, \beta_-)$ plane with the
value of $\Omega=0.7604$. This is a magnified region pointed by two arrows on
previous figure. Wide regions represents a second order structure. One can 
easily see a third order structure as thing lines.

\end{document}